# Microfluidic cell engineering on high-density microelectrode arrays for assessing structure-function relationships in living neuronal networks

Yuya Sato,[1,2] Hideaki Yamamoto,[1,a)] Hideyuki Kato,[3] Takashi Tanii,[4] Shigeo Sato,[1] and Ayumi Hirano-Iwata[1,2,5]

[1]*Research Institute of Electrical Communication, Tohoku University, 2-1-1 Katahira, Aoba-ku, Sendai 980-8577, Japan*

[2]*Graduate School of Biomedical Engineering, Tohoku University, 6-6 Aramaki-aza Aoba, Aoba-ku, Sendai 980-8579, Japan*

[3]*Faculty of Science and Technology, Oita University, 700 Dannoharu, Oita-shi, Oita 870-1192, Japan*

[4]*Faculty of Science and Engineering, Waseda University, 3-4-1 Ohkubo, Shinjuku-ku, Tokyo 169-8555, Japan*

[5]*Advanced Institute for Materials Research, Tohoku University, 2-1-1 Katahira, Aoba-ku, Sendai 980-8577, Japan*

[a)]*Author to whom correspondence should be addressed: hideaki.yamamoto.e3@tohoku.ac.jp*

**Abstract**

Neuronal networks in dissociated culture combined with cell engineering technology offer a pivotal platform to constructively explore the relationship between structure and function in living neuronal networks. Here, we fabricated defined neuronal networks possessing a modular architecture on high-density microelectrode arrays (HD-MEAs), a state-of-the-art electrophysiological tool for recording neural activity with high spatial and temporal resolutions. We first established a surface coating protocol using a cell-permissive hydrogel to stably attach polydimethylsiloxane microfluidic film on the HD-MEA. We then recorded the spontaneous neural activity of the engineered neuronal network, which revealed an important portrait of the engineered neuronal network—modular architecture enhances functional complexity by reducing the excessive neural correlation between spatially segregated modules. The results of this study highlight the impact of HD-MEA recordings combined with cell engineering technologies as a novel tool in neuroscience to constructively assess the structure-function relationships in neuronal networks.

**Main text**

Connectomics analyses provided anatomical information of the nervous system of animals at resolutions ranging from individual cells to brain regions.[1] The study has revealed several non-random properties such as the modular architecture as a structure that is evolutionarily conserved in the nervous system[2] and has given mechanistic insights into how network structure defines system functions in both normal and pathological brains.[3-5] While many studies have deciphered the structure-function relationships in the nervous system *in vivo*, recent advances in cell engineering technology using micropatterned proteins and microfluidic devices has enabled the use of cultured cells to study these relationships in a well-defined *in vitro* system.[6-15]

The two major technologies employed to record network activities in cultured neurons are fluorescence calcium imaging, which offer advantages in spatial resolution, and microelectrode arrays (MEA), which offer advantages in temporal resolution. Recently, high-density MEA (HD-MEA) technology has been developed, breaking the trade-off problem in spatial and temporal resolutions.[7,15-22] More precisely, recent HD-MEA devices offer spatial resolutions of over 3,000 electrodes $mm^{-2}$ with the electrode pitch below 20 μm and a temporal resolution below 100 μs.[15,20-22] The electrode pitch is comparable to the size of a neuronal cell body, and the temporal resolution is higher than a typical delay of synaptic transmission (~0.6 ms)[23]. Thus, the combination of the cell engineering and HD-MEA technologies provides a new framework to assess structure-function relationships in biological neuronal networks with unprecedented spatial and temporal resolutions.

Here, we fabricated neuronal networks possessing a modular architecture on HD-MEA using a polydimethylsiloxane (PDMS) microfluidic film and recorded their spontaneous activity at a resolution of 50 μs. Engineering neuronal networks of HD-MEA has been challenged by the surface topography originating in the passivation layer and the underlying electronics of the device,[17-18] which inhibited stable sealing of microfluidic devices.[15] We resolved this issue by coating the HD-MEA surface with a cell-permissive hydrogel smoothens the surface topography of the HD-MEA, enabling a gap-less adhesion of the PDMS microfluidic film to the HD-MEA. The high-temporal resolution of neural activity recording revealed an important portrait of neuronal networks that modular architecture enhances functional complexity by reducing the excessive neural correlation between spatially segregated modules. Our results highlight the impact of HD-MEA recordings combined with cell engineering technologies as a tool to assess the structure-function relationships in neuronal networks.

MaxOne HD-MEA chip (MaxWell Biosystems), bearing 26,400 electrodes with an interelectrode separation of 17.5 μm, was used in this study. The HD-MEA chip was first cleaned by air-plasma and 70% ethanol. The electrode area was then coated by a cell-permissive hydrogel and poly-D-lysine. The hydrogel layer was formed by drop casting 50 μL of collagen solution [1:1 mixture of type-I collagen solution (5 mg mL$^{-1}$, AteloCell IAC-50, Koken) and Neurobasal medium (Gibco)] on the electrode area of approximately 2.1 × 3.9 mm$^2$, aspirating excessive volume of the solution, and then inducing gelification of the remaining solution in an incubator and refrigerator.[24,25] PDMS microfluidic films for cell patterning were fabricated as detailed previously[12] and gently placed on the coated electrodes. The HD-MEA well was then filled with the neuronal plating medium [minimum essential medium + 10% fetal bovine serum], and primary neurons obtained from embryonic rat cortex were plated at a density of 5.3 × 10$^4$ cells/cm$^2$. The medium was changed to the Neurobasal medium [Neurobasal + 2% B-27 + 1% GlutaMAX-I] after 1 h, and half the medium was changed with the fresh Neurobasal medium twice every week. Spontaneous neural activity was recorded using the MaxLab Live software (MaxWell Biosystems) at a sampling frequency of 20 kHz. Action potentials were detected from high-pass filtered (>300 Hz) signal with a threshold of −5×SD. Local field potential (LFP) was extracted by down-sampling the recorded voltage traces to 1-ms resolutions and band-pass filtering at 1−100 Hz, and a negative peak deflection in the LFP (nLFP) was detected at each electrode by thresholding the LFP signal at −3×SD.[26]

A confocal micrograph (Keyence VK-X260) of the HD-MEA and its surface profile is shown in Fig. 1(a). The depth of the surface groove was approximately 1.5 µm, greater than the diameter of axons and dendrites.[27] We, thus, coated the HD-MEA surface with a collagen hydrogel layer with a thickness 0.44 ± 0.32 µm at its dried state (mean ± SD; $n = 16$ measurements from 8 samples in 3 preparations) [Fig. 1(b)]. With the reported swelling ratio of approximately five,[28] the hydrogel layer was sufficient to coat the surface topography of the HD-MEA. Primary neurons cultured on the HD-MEA with a PDMS microfluidic film placed on top are shown in Fig. 1(c). Without the hydrogel layer, non-specific growth of neurites was observed due to the space that remained at the base of the PDMS microfluidic film. Surface treatment with the hydrogel suppressed the non-specific neurite outgrowth, and the fraction of compliant networks, i.e., patterns without non-specific neurite growth, increased from 0.25 ($n = 8$ networks) to 0.67 ($n = 32$ networks) [Fig. 1(d)].

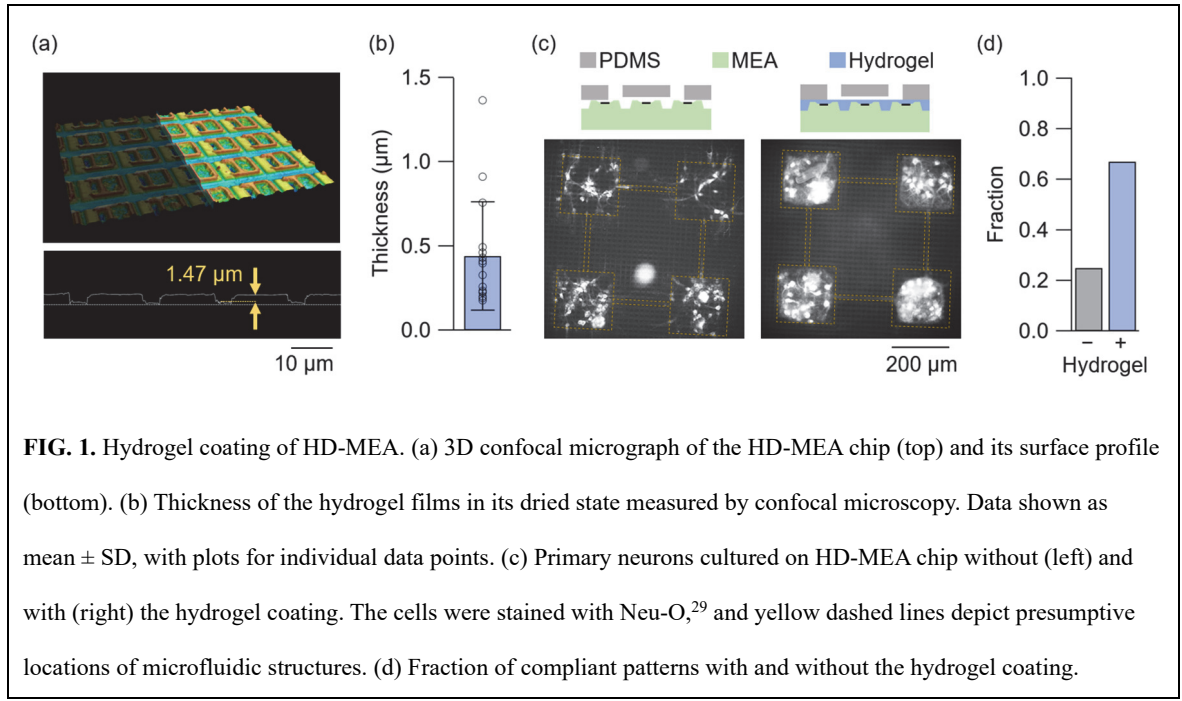

**FIG. 1.** Hydrogel coating of HD-MEA. (a) 3D confocal micrograph of the HD-MEA chip (top) and its surface profile (bottom). (b) Thickness of the hydrogel films in its dried state measured by confocal microscopy. Data shown as mean ± SD, with plots for individual data points. (c) Primary neurons cultured on HD-MEA chip without (left) and with (right) the hydrogel coating. The cells were stained with Neu-O,[29] and yellow dashed lines depict presumptive locations of microfluidic structures. (d) Fraction of compliant patterns with and without the hydrogel coating.

Insertion of the hydrogel layer increases the cell-electrode distance. We, therefore, assessed how much the hydrogel coating degrades the signal amplitude of the extracellular action potentials. Representative waveforms of extracellular action potentials recorded without and with the hydrogel layer are shown in Fig. 2(a) and 2(b), respectively. Presence of the hydrogel layer decreased the median signal amplitude by 29%, from 23.2 µV to 16.5 µV [Fig. 2(c)]. The signal amplitude, however, was still well above the noise level (<5 µV$_{rms}$) of the current MEA setup, securing the use of hydrogel coating as a novel approach to interface PDMS microfluidic device and HD-MEA.

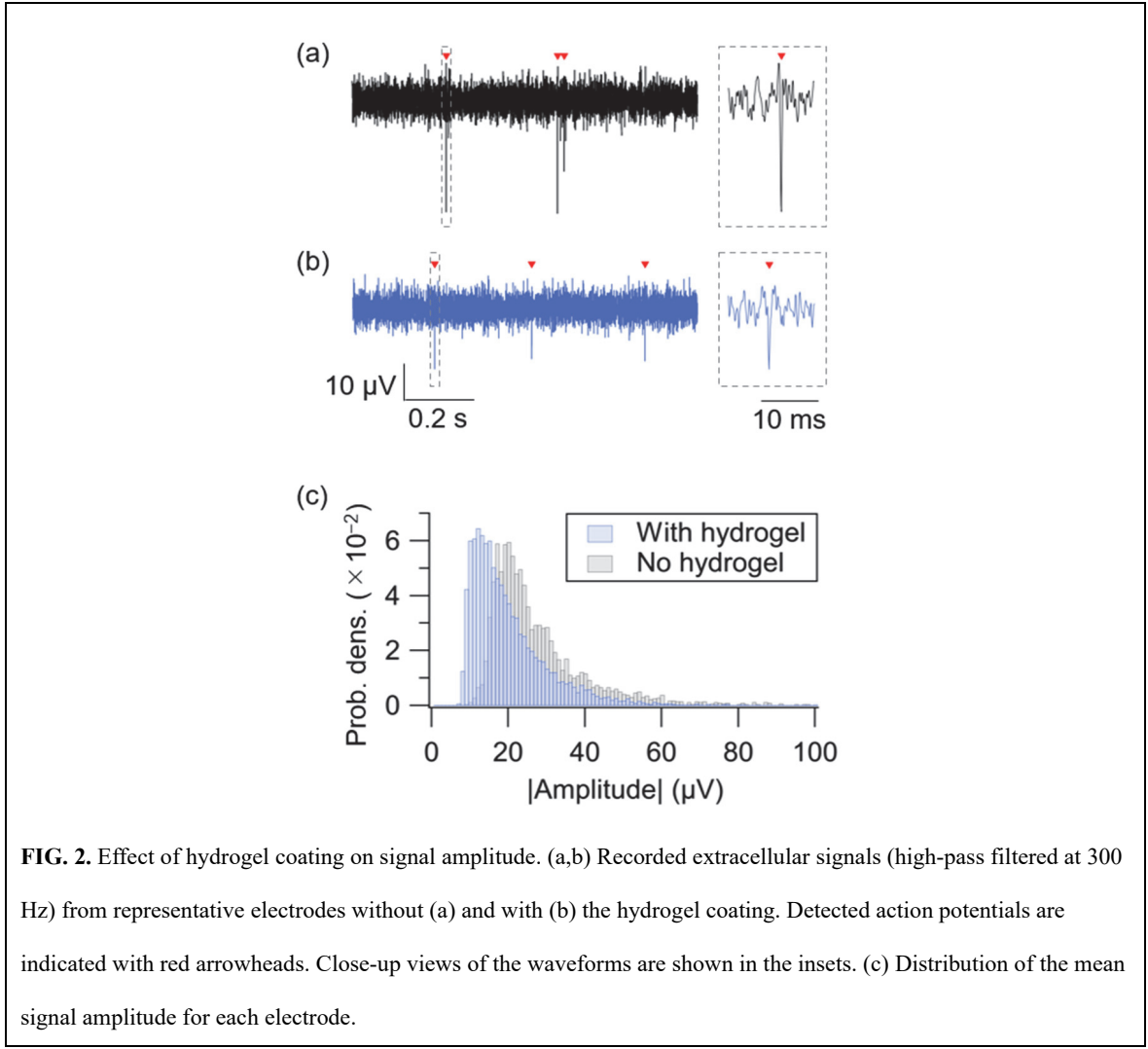

**FIG. 2.** Effect of hydrogel coating on signal amplitude. (a,b) Recorded extracellular signals (high-pass filtered at 300 Hz) from representative electrodes without (a) and with (b) the hydrogel coating. Detected action potentials are indicated with red arrowheads. Close-up views of the waveforms are shown in the insets. (c) Distribution of the mean signal amplitude for each electrode.

Figure 3 summarizes representative recordings of spontaneous neural activity from two types of micropatterned neuronal networks, i.e., a 'random' network and a 'modular' network. The random network was an isolated square of $400 \times 400$ μm$^2$, in which neurons grew uniformly and formed random connections [Fig. 3(a)]. The activity of the network was predominantly governed by the network bursts, i.e., a population activity that entrained a large fraction of the network [Fig. 3(b)].[30,31] This resulted in the pairwise correlation coefficient $r_{ij}$ to be high between a large fraction of electrode pairs [Fig. 3(c)]. Here, $r_{ij}$ was calculated as $r_{ij} =$

$\frac{\text{Cov}(n_i, n_j)}{\sqrt{\text{Var}(n_i)\text{Var}(n_j)}}$, where $i$ and $j$ are electrode indices, $n_i = n_i(t)$ is the spike train of electrode $i$, and Cov and Var are the covariance and variance over the entire time bins, respectively. The bin width was set to 50 ms, and $n_i(t)$ was one if the electrode detected more than one spikes in the $t$-th time bin, and zero otherwise.

Modular networks, in contrast, exhibited a richer repertoire of population activity. Modular networks comprised four squares, or modules, of 200 × 200 μm² connected by microchannels with widths of 6.7 ± 0.79 μm ($n$ = 12) [Fig. 3(d)]. The microchannel allowed axons and dendrites of a fraction of neurons to project to neighboring modules and form functional couplings between them.[12] While activity of the neurons in the same module was strongly correlated, the interaction of neurons in separate modules were weaker, leading to probabilistic coherence [Fig. 3(e) and 3(f)].

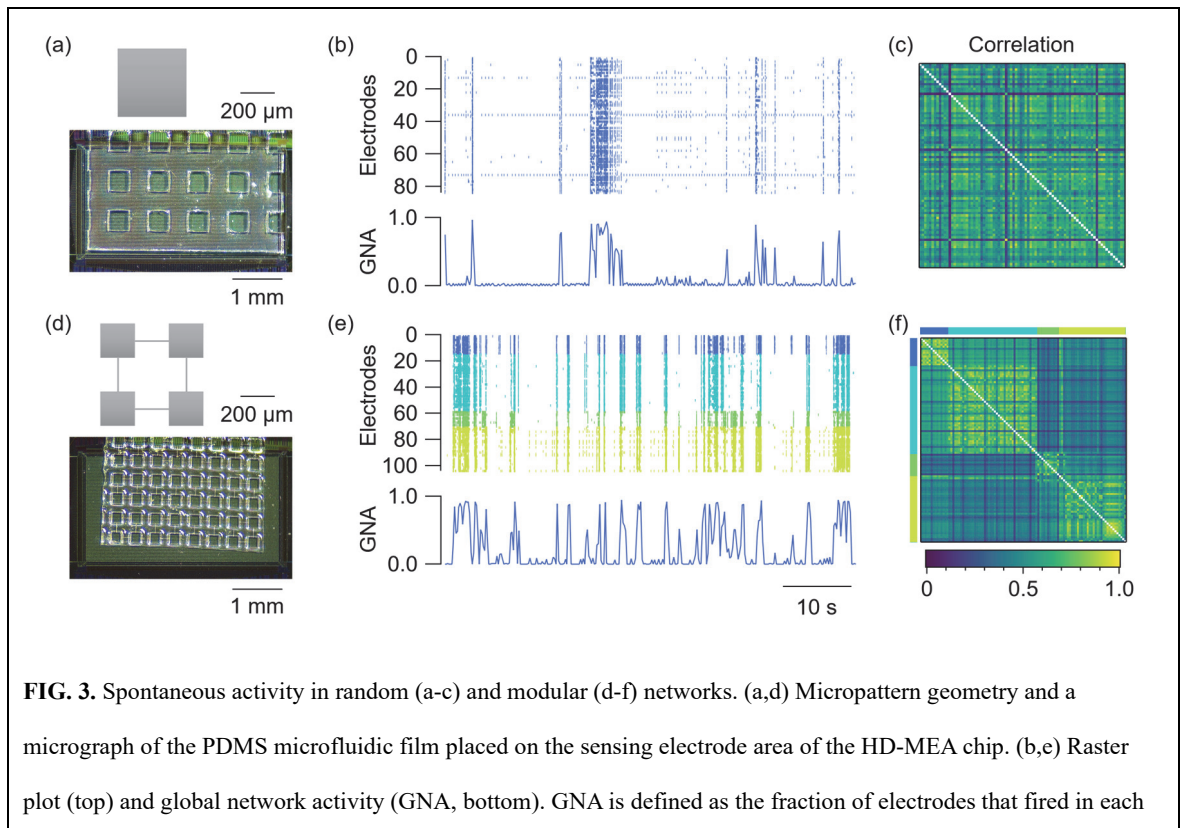

**FIG. 3.** Spontaneous activity in random (a-c) and modular (d-f) networks. (a,d) Micropattern geometry and a micrograph of the PDMS microfluidic film placed on the sensing electrode area of the HD-MEA chip. (b,e) Raster plot (top) and global network activity (GNA, bottom). GNA is defined as the fraction of electrodes that fired in each time bin. The width of the time bin was set to 200 ms. (c,f) Correlation coefficient matrix. For the modular network, electrodes belonging to separate modules are assigned different colors in (e,f).

To statistically evaluate the impact of network structure on the dynamics of spontaneous activity, we further assessed functional complexity, functional modularity, and the statistics of $r_{ij}$ and neural avalanches in the random and modular networks. Functional complexity $C$ is a measure of integration-segregation balance important for proper functioning of the nervous system, with $C = 1$ and $0$ indicating maximally and minimally balanced states, respectively.[10,32] It is defined as:

$$C = 1 - \frac{1}{C_m} \sum_{\mu=1}^{m} \left| p_\mu(r_{ij}) - \frac{1}{m} \right| \qquad (1)$$

where $p_\mu(r_{ij})$ is the probability of $r_{ij}$ to be in the $\mu$-th bin, $C_m = 2(m-1)/m$ is a normalization factor, and $m = 20$. For the statistical analysis, electrodes with a mean firing rate below 0.5 Hz were excluded from the analysis as non-active electrodes, and networks with more than 20 active electrodes were used. Previous studies using calcium imaging have revealed that patterning cortical neurons in modular architecture broadens the distribution of $r_{ij}$ and increases the value of $C$.[10,12] Analyses of the spontaneous neural activity recorded at 10–14 DIV revealed that the value of $C$ was significantly higher in the modular network than in the random network, confirming the previous observations [Fig. 4(a); $p < 0.01$, $n = 14$ for both random and modular networks; two-tailed $t$-test].

The degree of modularization in the functional connectivity of the networks was then quantified by calculating modularity $Q$ of the correlation matrix:[33]

$$Q = \frac{1}{2E} \sum_{i=1}^{N} \sum_{j=1}^{N} \left( A_{ij} - \frac{k_i k_j}{2E} \right) \delta_{m_i m_j} \qquad (2)$$

where $A = [A_{ij}]$ is a binarized correlation matrix generated by thresholding $r_{ij}$ at 0.7, $2E =$

$\sum_{ij} A_{ij}$ is the total number of edges ($A_{ij}$ = 1) in the matrix, $k_i = \sum_j A_{ij}$ is the node degree of $i$, $\delta_{m_i m_j}$ is the Kronecker delta function which is equal to one if nodes $i$ and $j$ belong to the same module ($m_i = m_j$) and zero otherwise. A positive value of $Q$ indicates the presence of modular structure with a maximum of $Q$ = 1. In contrast, $Q$ = 0 if the matrix lacks modular structure and is random. The analysis revealed that the functional connectivity is more modularized in the modular network [Fig. 4(b); $p < 0.01$, $n$ = 13 and 11 for random and modular networks, respectively; two-tailed $t$-test], underscoring that structural confinement induced functional modularization in the cultured neuronal network.

We next took the advantage of high temporal resolution of MEA recordings to assess how the mean correlation coefficient $R = \sum_{ij} r_{ij}/(N^2 - N)$ of modular networks depends on the temporal bin width [Fig. 4(c)]. To this end, we classified $R$ into three categories: (1) $R_{in}$, the mean value of $r_{ij}$ evaluated within each module; (2) $R_{neib}$, the mean evaluated across neighboring modules; and (3) $R_{non-neib}$, the mean evaluated across non-neighboring modules. As a general trend, the values of $R$ were dependent on the bin width of the spike train and increased when larger bin was used. Comparison at constant bin width revealed that of $R_{in}$ was larger than $R_{neib}$ and $R_{non-neib}$ at any bin width, indicating that intramodular correlation was greater than intermodular ones. The difference between $R_{neib}$ and $R_{non-neib}$ was smaller than that against $R_{in}$. However, $R_{non-neib}$ was found to be significantly smaller than $R_{neib}$ for the bin width between 5 ms and 20 ms ($p < 0.05$; $n$ = 14 networks; two-tailed $t$-test). The lower value of $R_{non-neib}$ is expected as non-neighboring modules are spatially more distanced than neighboring modules, increasing the signal delay via signal propagation and synaptic transmission. Importantly, such analysis was impossible with low-temporal resolution recordings of neural activity and highlights a novel potential of MEA recordings in the assessment of structure-function relationships in living neuronal networks.

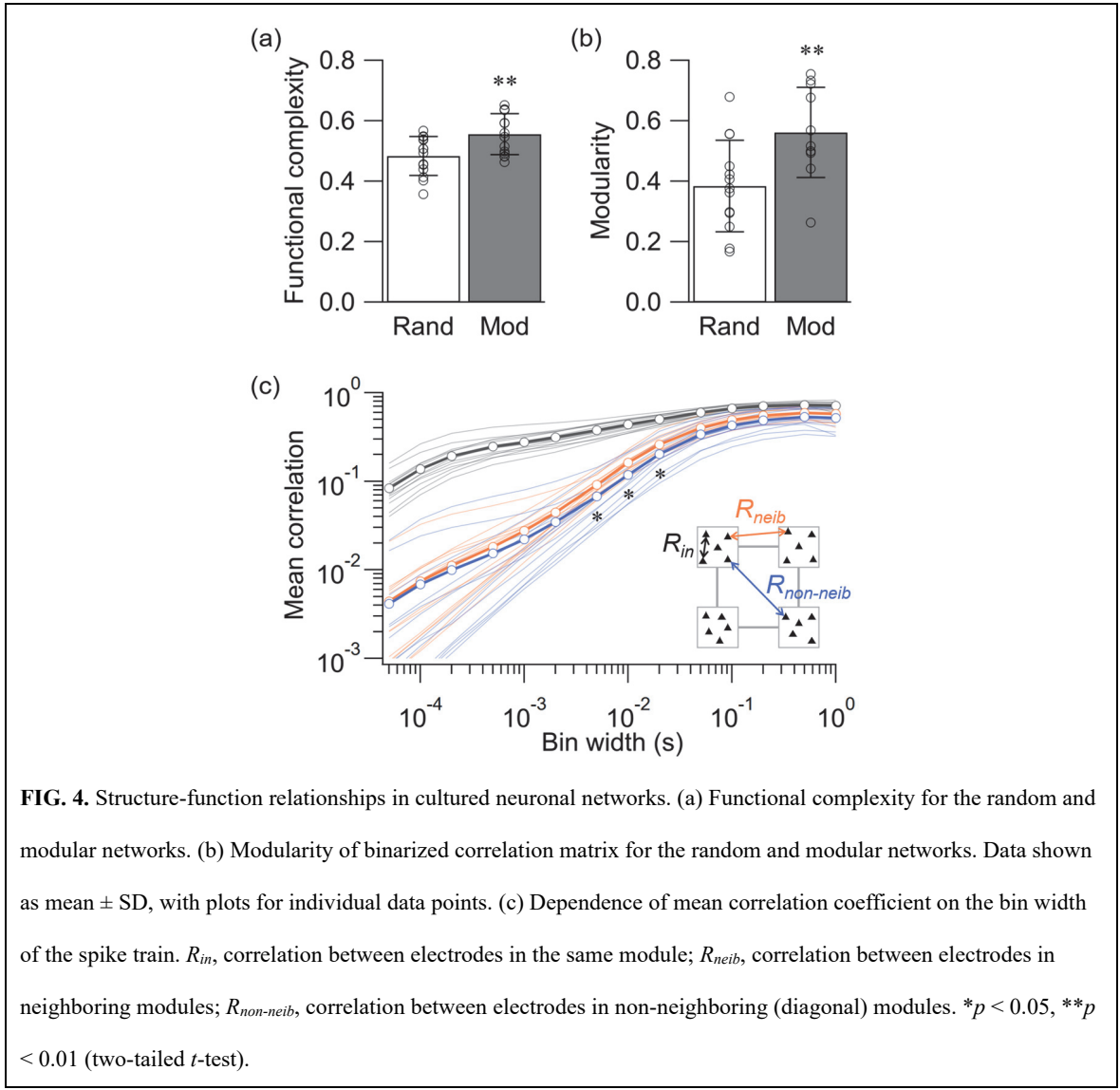

**FIG. 4.** Structure-function relationships in cultured neuronal networks. (a) Functional complexity for the random and modular networks. (b) Modularity of binarized correlation matrix for the random and modular networks. Data shown as mean ± SD, with plots for individual data points. (c) Dependence of mean correlation coefficient on the bin width of the spike train. $R_{in}$, correlation between electrodes in the same module; $R_{neib}$, correlation between electrodes in neighboring modules; $R_{non\text{-}neib}$, correlation between electrodes in non-neighboring (diagonal) modules. $*p < 0.05$, $**p < 0.01$ (two-tailed $t$-test).

Finally, we compared the spatiotemporal structure of the spontaneous neural dynamics in random and modular networks by analyzing the neural avalanches in LFP signals [Fig. 5(a)]. Neural avalanche is defined as a sequence of time bins during which an nLFP was detected in at least one electrode. An nLFP reflects the summation of local inward currents, and its analysis uncovers whether the system operates near a critical point, a point at which an activity stably propagates and maximizes information transmission within a network.[26,34] Accumulating experimental evidence further support the hypothesis that the mammalian cortex self-organizes into a critical state.[35,36] We thus analyzed the nLFP statistics in each network [Fig. 5(b)] by calculating the branching parameter, as well as the probability distribution of avalanche sizes and durations.

Stability of the propagation can be measured by evaluating the branching parameter σ*:[37]

$$\sigma^* = \langle \sigma_i^* \rangle = \langle \text{round}\left(\frac{n_{i+1}}{n_i}\right)\rangle \quad (3)$$

where σ*$_i$ is an estimate for the *i*-th bin in all avalanches of a recording, $n_i$ is the number of active electrodes at *i*-th bin, round is the rounding operation to the nearest integer, and <> is the average over all *i*'s. σ*$_i$ was not calculated for $n_i$ = 0. Evaluation of σ* revealed that value was slightly above 1 for both random and modular networks, suggesting a near-critical state with a tendency toward supercriticality [Fig. 5(c)]. When the network is in a critical state, neuronal activity neither increases nor decreases in avalanches and thus stably propagates in time and space. When the network is supercritical, the activity tends to expand once initiated.

The difference between the random and modular architectures was most evident in the avalanche size distribution. For a neuronal network operating near a critical point, distributions of avalanche sizes *S* and avalanche duration *T* are scale-invariant and obey a power law such that $p(S) \propto S^{-\tau}$ and $p(T) \propto T^{-\alpha}$, where τ and α are the power-law exponents. The distribution of avalanche sizes and durations averaged over the samples are shown in Fig. 5(d) and 5(e), respectively, for both the random and modular networks. The avalanche size distribution for the random network exhibited a deviation from the power law as a small peak near *S* = 20 to 40, which is an indication of supercritical dynamics. A peak was less prominent and shifted towards a smaller value of *S* in the recordings from the modular networks, which is most probably due to the segregation of the neuronal network in subpopulations. This result suggests that fabrication of structured networks with a larger number of modules may help to realize engineered neuronal networks with dynamics near criticality, along with a proper balancing of excitation and inhibition that develops with the days in culture and is a strong control parameter for avalanche dynamics.[35,36,38]

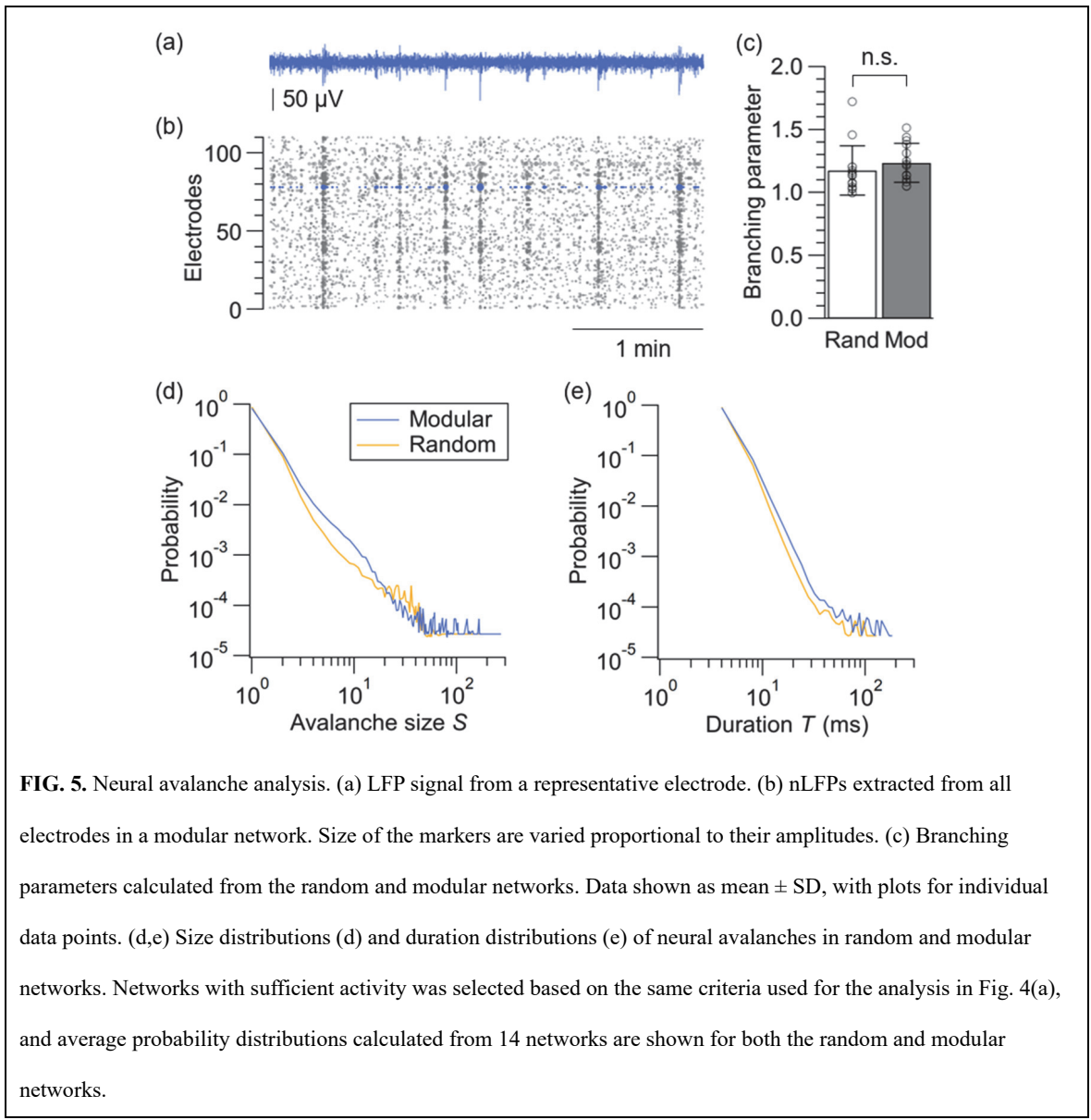

**FIG. 5.** Neural avalanche analysis. (a) LFP signal from a representative electrode. (b) nLFPs extracted from all electrodes in a modular network. Size of the markers are varied proportional to their amplitudes. (c) Branching parameters calculated from the random and modular networks. Data shown as mean ± SD, with plots for individual data points. (d,e) Size distributions (d) and duration distributions (e) of neural avalanches in random and modular networks. Networks with sufficient activity was selected based on the same criteria used for the analysis in Fig. 4(a), and average probability distributions calculated from 14 networks are shown for both the random and modular networks.

Although the cell-electrode configuration in the current experiment seemingly contradicts with the classic point-contact model of extracellular recordings, non-contact recordings of extracellular action potentials have previously been demonstrated in cardiomyocytes,[39] and their mechanism can be described by the volume conductor theory.[19] Increased cell-electrode distance, however, inevitably decreases the signal amplitude and may impede applications in recordings that demand higher signal-to-noise ratio. In such a case, other approaches for sealing microfluidic devices, such as the one that uses thin layer of PDMS gel to glue microfluidic devices to the electrodes,[15] may need to be considered. Otherwise, our proposed method is advantageous in enabling the use of sub-10 μm microchannels, which was challenging in the PDMS-gluing approach. Our method also allows repeated use of an HD-MEA chip for at least three times and is economical.

In conclusion, we established a protocol to interface electrode surface with microfluidic devices for controlling the structure of cultured neuronal networks on HD-MEA. The hydrogel coating of device surface inevitably decreased the signal amplitude of extracellular action potentials, but the signal was large enough to be detected from microelectrodes. A small neuronal network bearing modular architecture was fabricated and its dynamics was assessed based on the statistics of correlation coefficients and neural avalanches. High-temporal resolution of the MEA, along with its potential to record LFPs, revealed novel aspect of the engineered networks, which was inaccessible with fluorescence calcium imaging. Cell engineering based on microfluidics has become an indispensable technology for studying structure-function relationships and modeling neuronal network functions *in vitro*. In addition to network modularity, precise control of axon orientation[40] and even dendritic spines[41] has been demonstrated. Combination of microfluidics-based neuroengineering with state-of-the-art MEA devices will open new application of *in vitro* systems as a tool in fundamental neuroscience and pharmacology.

**Acknowledgement**

This study was supported by the Cooperative Research Project Program of Research Institute of Electrical Communication (RIEC), Tohoku University, MEXT KAKENHI Grant-in-Aid for Transformative Research Areas (B) "Multicellular Neurobiocomputing" (21H05163, 21H05164), JSPS KAKENHI (18H03325, 20H00594, 20H02194, 21K12050), JST-PRESTO (JMPJPR18MB), and JST-CREST (JPMJCR19K3). This research has been partly carried out at the Laboratory for Nanoeletcronics and Spintronics, RIEC, Tohoku University and the Fundamental Technology Center, RIEC, Tohoku University.

**Author Declarations**

**Conflict of Interest.** The authors have no conflicts of interest to declare.

**Ethics Approval.** All experimental procedures regarding the use of animals were approved by the Tohoku University Center for Laboratory Animal Research, Tohoku University (approval number: 2020AmA-001).

**Data Availability**

The data that support the findings of this study are available from the corresponding author upon reasonable request.

**References**

1. R. Gao et al., Science **363**, eaau8302 (2019).

2. M. P. van den Heuvel, E. T. Bullmore, and O. Sporns, Trends Cogn. Sci. **20**, 345 (2016).

3. C. W. Lynn and D. S. Bassett, Nat. Rev. Phys. **1**, 318 (2019).

4. L. E. Suárez, B. A. Richards, G. Lajoie, and B. Misic, Nat. Mach. Intell. **3**, 771 (2021).

5. M. P. van den Heuvel and O. Sporns, Nat. Rev. Neurosci. **20**, 435 (2019).

6. O. Feinerman, A. Rotem, and E. Moses, Nat. Phys. **4**, 967 (2008).

7. M. K. Lewandowska, D. J. Bakkum, S. B. Rompani, and A. Hierlemann, PLoS ONE **10**, e0118514 (2015).

8. H. Yamamoto, R. Matsumura, H. Takaoki, S. Katsurabayashi, A. Hirano-Iwata, and M. Niwano, Appl. Phys. Lett. **109**, 043703 (2016).

9. J. Albers and A. Offenhäusser, Front. Bioeng. Biotechnol. **4**, 46 (2016).

10. H. Yamamoto, S. Moriya, K. Ide, T. Hayakawa, H. Akima, S. Sato, S. Kubota, T. Tanii, M. Niwano, S. Teller, J. Soriano, and A. Hirano-Iwata, Sci. Adv. **4**, eaau4914 (2018).

11. C. Forró, G. Thompson-Steckel, S. Weaver, S. Weydert, S. Ihle, H. Dermutz, M. J. Aebersold, R. Pilz, L. Demkó, and J. Vörös, Biosens. Bioelectron. **122**, 75 (2018).

12. T. Takemuro, H. Yamamoto, S. Sato, and A. Hirano-Iwata, Jpn. J. Appl. Phys. **59**, 117001 (2020).

13. S. J. Ihle, S. Girardin, T. Felder, T. Ruff, J. Hengsteler, J. Duru, S. Weaver, C. Forró, and J. Vörös, Biosens. Bioelectron. **201**, 113896 (2022).

14. N. Hong, Y. Nam, Mol. Cells **45**, 76 (2022).

15. J. Duru, J. Kuchler, S. J. Ihle, C. Forró, A. Bernardi, S. Girardin, J. Hengsteler, S. Wheeler, J. Vörös, and T. Ruff, Front. Neurosci. **16**, 829884 (2022).

16. L. Berdondini, K. Imfeld, A. Maccione, M. Tedesco, S. Neukom, M. Koudelka-Hep, and S.

Martinoia, Lab Chip **9**, 2644 (2009).

17. U. Frey, J. Sedivy, F. Heer, R. Pedron, M. Ballini, J. Mueller, D. Bakkum, S. Hafizovic, F. D. Faraci, F. Greve, K.-U. Kirstein, and A. Hierlemann, IEEE J. Solid-State Circ. **45**, 467 (2010).

18. A. Hierlemann, U. Frey, S. Hafizovic, and F. Heer, Proc. IEEE **99**, 252 (2011).

19. M. E. J. Obien, K. Deligkaris, T. Bullmann, D. J. Bakkum, and U. Frey, Front. Neurosci. **8**, 423 (2015).

20. X. Yuan et al., Nat. Commun. **11**, 4854 (2020).

21. N. A. Steinmetz et al., Science **372**, eabf4588 (2021).

22. K. Shimba, T. Asahina, K. Sakai, K. Kotani, and Y. Jimbo, Front. Neurosci. **16**, 854637 (2022).

23. J. E. Lisman, S. Raghavachari, and R. W. Tsien, *Nat. Rev. Neurosci.* **8**, 597 (2007).

24. T. Takezawa, K. Ozaki, A. Nitani, C. Takabayashi, and T. Shimo-oka, Cell Transplant. **13**, 463 (2004).

25. R. Matsumura, H. Yamamoto, M. Niwano, and A. Hirano-Iwata, Appl. Phys. Lett. **108**, 023701 (2016).

26. J. M. Beggs and D. Plenz, J. Neurosci. **23**, 11167 (2003).

27. W. P. Bartlett and G. A. Banker, J. Neurosci. **4**, 1944 (1984).

28. C.-L. Kim and D.-E. Kim, Sci. Rep. **6**, 20563 (2016).

29. J. C. Er et al., Angew. Chem. Int. Ed. **54**, 2442 (2015).

30. J. G. Orlandi, J. Soriano, E. Alvarez-Lacalle, S. Teller, and J. Casademunt, Nat. Phys. **9**, 582 (2013).

31. H. Yamamoto, S. Kubota, Y. Chida, M. Morita, S. Moriya, H. Akima, S. Sato, A. Hirano-Iwata, T. Tanii, and M. Niwano, Phys. Rev. E **94**, 012407 (2016).

32. G. Zamora-López, Y. Chen, G. Deco, M. L. Kringelbach, and C. Zhou, Sci. Rep. **6**, 38424 (2016).

33. M. E. J. Newman, *Proc. Natl. Acad. Sci., U.S.A.* **103**, 8577 (2006).

34. D. Plenz and T. C. Thiagarajan, Trend. Neurosci. **30**, 101 (2007).

35. M. A. Muñoz, Rev. Mod. Phys. **90**, 031001 (2018).

36. D. Plenz, T. L. Ribeiro, S. R. Miller, P. A. Kells, A. Vakili, and E. L. Capek, Front. Phys. **9**, 639389 (2021).

37. V. Priesemann, M. Wibral, M. Valderrama, R. Pröpper, M. Le Van Quyen, T. Geisel, J. Triesch, D. Nikolić, and M. H. J. Munk, Front. Syst. Neurosci. **8**, 108 (2014).

38. Y. Yada, T. Mita, A. Sanada, R. Yano, R. Kanzaki, D. J. Bakkum, A. Hierlemann, and H. Takahashi, Neuroscience **343**, 55 (2017).

39. T. Sharf, P. K. Hansma, M. A. Hari, and K. S. Kosik, Lab Chip **19**, 1448 (2019).

40. J.-M. Peyrin, B. Deleglise, L. Saias, M. Vignes, P. Gougis, S. Magnifico, S. Betuing, M. Pietri, J. Caboche, P. Vanhoutte, J.-L. Viovy, and B. Brugg, Lab Chip **11**, 3663 (2011).

41. J. C. Mateus, S. Weaver, D. van Swaay, A. F. Renz, J. Hengsteler, P. Aguiar, and J. Vörös, ACS Nano **16**, 5731 (2022).